\documentclass[aip,twocolumn,reprint]{revtex4}

\usepackage{graphicx}
\usepackage{dcolumn}
\usepackage{bm}

\begin{document}
{accepted in APL (2011)}

\title{Oriented polaritons in strongly-coupled asymmetric double quantum well microcavities} 



\author{Gabriel Christmann}
\email{gprmc2@cam.ac.uk}
\affiliation{Cavendish Laboratory, University of Cambridge, Cambridge CB3 0HE, United Kingdom.}
\author{Alexis Askitopoulos}
\affiliation{Department of Materials Science and Technology, University of Crete, P.O. Box 2208, 71003 Heraklion, Greece.}
\author{George Deligeorgis}
\author{Zacharias Hatzopoulos}
\affiliation{Microelectronics Research Group, IESL-FORTH, P.O. Box 1385, 71110 Heraklion, Greece.}
\author{Simeon I. Tsintzos}
\author{Pavlos G. Savvidis}
\affiliation{Department of Materials Science and Technology, University of Crete, P.O. Box 2208, 71003 Heraklion, Greece.}
\affiliation{Microelectronics Research Group, IESL-FORTH, P.O. Box 1385, 71110 Heraklion, Greece.}
\author{Jeremy J. Baumberg}
\affiliation{Cavendish Laboratory, University of Cambridge, Cambridge CB3 0HE, United Kingdom.}


\date{\today}

\begin{abstract}
Replacing independent single quantum wells inside a strongly-coupled semiconductor microcavity with double quantum wells produces a special type of polariton. Using asymmetric double quantum wells in devices processed into mesas allows the alignment of the electron levels to be voltage-tuned. At the resonant electronic tunnelling condition, we demonstrate that `oriented polaritons' are formed, which possess greatly enhanced dipole moments. Since the polariton-polariton scattering rate depends on this dipole moment, such devices could reach polariton lasing, condensation and optical nonlinearities at much lower threshold powers.
\end{abstract}


\maketitle 


Semiconductor microcavities (MCs) in the strong coupling regime (SCR) have shown great potential for the realization of optoelectronic devices. In this context, Imamo\u{g}lu and coworkers proposed in 1996 the concept of the polariton laser in which a MC in the SCR can emit coherent directional light, just like a vertical cavity surface emitting laser (VCSEL), but below the inversion threshold.\cite{Imamoglu96} Working devices were subsequently realized with several semiconductors systems: II-tellurides\cite{Dang98} and III-arsenides\cite{Bajoni08} at cryogenic temperatures, and even at room temperature with III-nitrides.\cite{Christopoulos07,Christmann08} Fabrication of fully functional devices is becoming even more realistic since electrically-injected structures exhibiting polariton light emission have been realized.\cite{Tsintzos08,Bajoni08led,Khalifa08}. The enormous polariton parametric amplification enables controllable nonlinear optical elements,\cite{Savvidis00,Christmann10} while more recently Bose-Einstein condensation\cite{Kasprzak06,Balili07} which is currently pushed towards room temperature with III-nitrides,\cite{Baumberg08,Levrat10} opens perspectives for the realization of coherent superfluid devices.\cite{Lagoudakis10}

Nevertheless currently the threshold of these devices is limited by the rate of polariton-polariton scattering. This is illustrated by the large number of quantum wells (QWs) generally inserted in the MCs for polariton lasing/condensation,\cite{Dang98,Bajoni08,Christmann08} compared to equivalent VCSEL structures.\cite{Jewell91,Kao08} This design is needed to allow high enough polariton densities for polariton-polariton scattering to become faster than polariton decay. Reducing the decay rate by using high-$Q$ cavities has led to polariton localisation in the photonic disorder potential. The desire to reduce the minimum threshold for polariton lasing/condensation has thus led to various proposals to enhance polariton relaxation. One suggestion has been to use scattering by free electrons,\cite{Malpuech02,Lagoudakis03} however to date this has not been effective.

In this letter we demonstrate an innovative MC design in which the active region is composed of asymmetric double quantum wells (ADQWs). By applying an electric field to the structure it is possible to bring the electron levels of neighboring quantum wells into resonance. In this resonant condition the spatially direct and indirect excitons become coupled so they share the strong oscillator strength of the direct constituent and the strong dipole moment in the growth direction of the indirect one.\cite{Ciuti98} This configuration should favour polariton-polariton interactions and is therefore very promising for threshold reduction of nonlinear effects in strong-coupling MCs. In addition, the ability to tune the polaritons by applied voltage offers a simple way to make polariton waveguides and devices. This concept thus combines ideas of controlling indirect excitons in bare QWs\cite{Butovpapers} with traditional strong coupling microcavities.

The sample used to demonstrate {\it oriented polaritons} is a strongly-coupled MC made of a $5\lambda/2$ undoped cavity containing four sets of In$_{0.1}$Ga$_{0.9}$As/GaAs/In$_{0.08}$Ga$_{0.92}$As (10 nm/4 nm/10 nm) ADQWs, placed at the antinodes of the electric field [Fig. 1(a)]. The cavity is formed from top (17-pair, $p$-doped) and bottom (21-pair $n$-doped) GaAs/AlAs distributed Bragg reflectors (DBRs), thus forming a $p-i-n$ junction. A $\Omega_{VRS}\sim 5.6$ meV Rabi splitting is measured at $10$ K in this structure. Polariton LEDs are processed into 400 $\mu$m diameter mesas with a ring-shaped Ti/Pt electrode deposited after a second etch step to contact the lower $p$-layers, improving the series resistance (details in \cite{Tsintzos08,Tsintzos09}).

In the ADQW, the lower energy excitons in the left quantum well (LQW, 10\% In) couple to the cavity mode while the higher energy exciton in the right quantum well (RQW, 8\% In) is blue shifted well out of resonance. Applying reverse bias decreases the electron levels of the RQW, producing a tunnelling resonance when it matches the LQW $n$=1 electron energy. At resonance, the electron wavefunctions will be spread between both wells [Fig. 1(b)] whereas the LQW hole remains strongly confined, thus creating a significant dipole moment.

\begin{figure}[htbp]
\centerline{\includegraphics[width=\columnwidth]{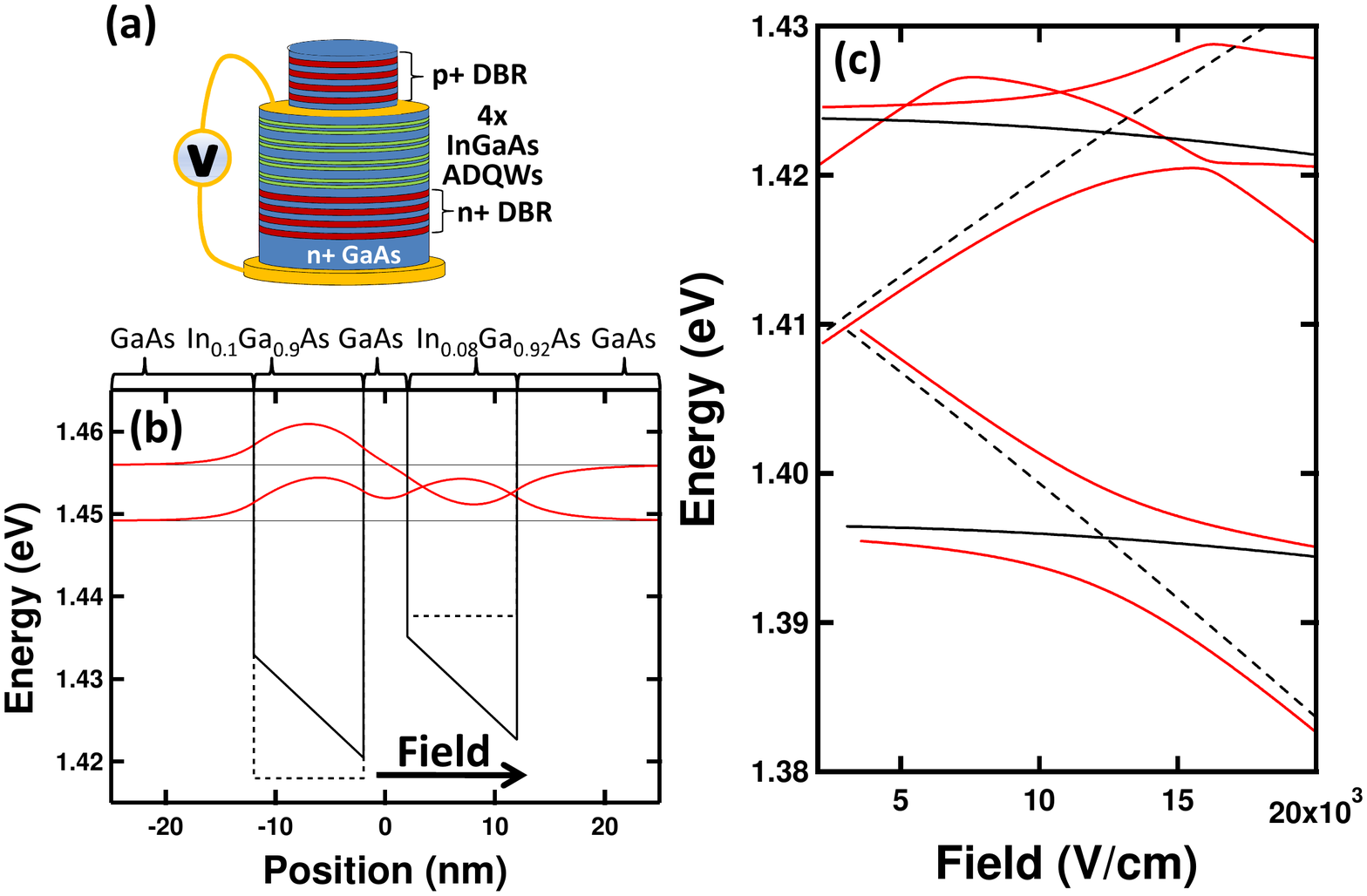}}
\caption{(color online) (a) ADQW polariton mesa structure. (b) Conduction bands of one ADQW period (black line at resonance, dashed at flat-band), and corresponding electron wavefunctions [red (dark gray) lines]. (c) Calculated transition energies for the ADQW structure [red (dark gray) lines], and for corresponding single QWs (black lines).}
\end{figure}

To understand the polaritons we compare energies of the direct (DX) and indirect (IX) excitons of the LQW coupled to the cavity mode. The indirect transition excites electrons from the LQW heavy-hole level to the RQW electron level. Solving the Schr\"odinger equation using complex Airy functions\cite{Ahn86} and suitable material parameters\cite{Vurgaftman01} yields complex eigenvalues where the real part is the quasiconfined energy level and the imaginary part is related to the carrier escape time. Compared to weak field tuning\cite{Skolnickpaperaround1998} of single QWs used in previous microcavities (black lines), a new level structure is revealed [Fig. 1(c)]. The tunnelling resonance at $12.5$ kV/cm between direct and indirect excitons leads to an anticrossing with tunnelling-induced splitting, $\hbar J$=5 meV. We concentrate here on the lowest states corresponding to the lower polariton branch only, although we also observe strong coupling with RQW excitons. The lifetime of the levels (not shown) is also obtained from the model. When these times become smaller than the Rabi period $T_{VRS}=2\pi/\Omega_{VRS}\sim 740$ fs, the strong coupling is lost. From our simulations this only occurs for $F>38$ kV/cm, hence the tunnelling resonance does not destroy the strong coupling.

\begin{figure}[htb]
\centerline{\includegraphics[width=\columnwidth]{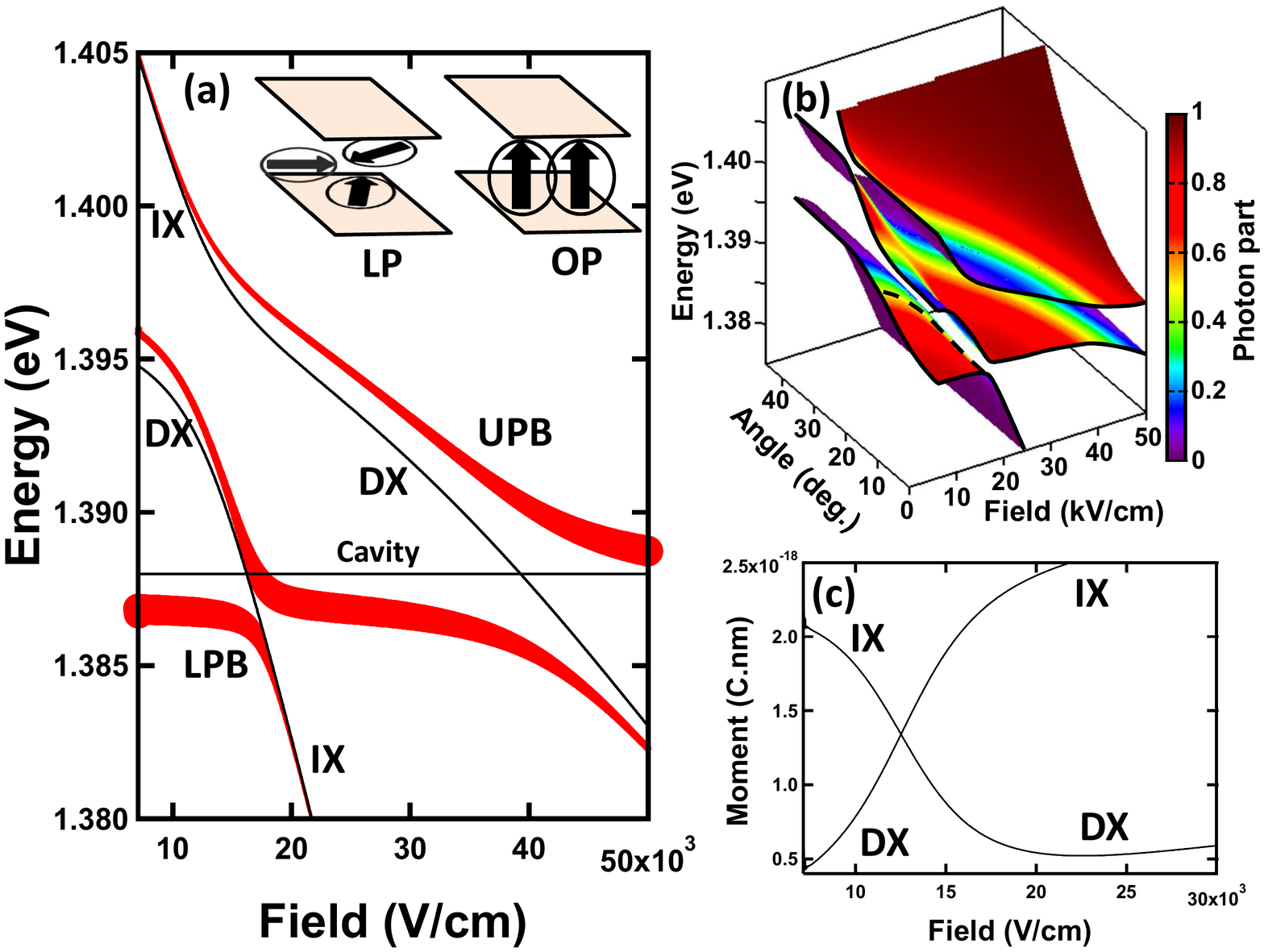}}
\caption{(color online) (a) Calculated polariton dispersion curves \textit{vs} electric field [red (dark gray) lines]. Line thickness is proportional to the photon fraction. Black lines are uncoupled modes (cavity mode $C$ and two excitons $DX$ and $IX$). Inset: dipole orientation for normal MC and for oriented polarions (OPs). (b) Surface plot of the polariton modes \textit{vs} electric field and incident angle. Surface color gives photon fraction. Black lines are guide to the eye. (c) Calculated dipole moment corresponding to $DX$ and $IX$, for single electron-hole pair.}
\end{figure}

From the corresponding wavefunctions produced by the model, the overlaps between electron and hole wavefunctions, which are proportional to the oscillator strength, are calculated for each level as a function of the applied field (not shown).\cite{noteWF} These coupling strengths are inserted in a standard $3\times 3$ Hamiltonian model to describe the SCR between the two excitons (direct and indirect) of the LQW and the cavity mode. The three resulting dispersion curves, lower, middle and upper polariton branches (LPB, MPB and UPB), are shown in Fig. 2(a), where the thickness of the lines is proportional to the photon fraction (and hence the out-coupling to free space). The effect of the tunnelling resonance is clearly observed on the LPB as an extra anticrossing. Surprisingly, by controlling the cavity detuning, the electric field of this anticrossing shifts relative to the bare electron tunnelling resonance. The important consequence of this observation is that the field at which this oriented polariton anticrossing occurs will be angle dependent [Fig. 2(b), dashed line]. This point opens interesting potential for the study of angle-resonant parametric scattering,\cite{Savvidis00} as pump, signal and idler will meet this anticrossing at different electric fields. 

Of key significance is the net polariton dipole moment at the resonance [Fig. 2(c)] induced by the electron spreading between QWs. Since polariton scattering depends on the exciton dipole-dipole coupling, this will now be greatly amplified compared to normal MCs [Fig. 2(a), inset]. The extent of the alignment depends on temperature, disappearing only for $T > \hbar J/k_B \simeq 60$ K here. Since the tunnelling rate is controlled exponentially by the barrier thickness, higher temperature operation is possible.
The oriented polaritons, 
\begin{equation}
\left| OP \right\rangle_{\pm} = a \left\{ \left| DX \right\rangle \pm \left| IX \right\rangle \right\} + c \left| C \right\rangle	\nonumber
\end{equation}
both have the same dipole moment, although the upper level has less penetration into the barrier [Fig. 1(b)]. 

Evidence for oriented polaritons is indeed observed in these devices. Broadband 150fs pulses of a Ti:S laser are used to record 40 K reflectivity spectra while changing the bias from $1.5$ V to $-2$ V [Fig. 3(a)]. The lower polariton (LP) mode is tracked as its reflectivity dip redshifts revealing a clear anticrossing. At large negative bias the features become weak and beyond this a non-dispersing broad reflectivity dip is seen at a slightly higher energy than the original LP mode. To compare this data to simulations, the applied bias is converted into internal electric field, $F=\left(V_g-V_b\right)/L_{i}$, with built-in potential $V_g$=1.52V and undoped $i$-thickness $L_i$=670 nm. In practice the effective field shifts due to series resistances of the DBRs and to compensating electric fields created by the non-zero excitonic dipole moment under illumination.\cite{Bajoni08bistable,Christmann10}. We thus perform measurements at low incident power with internal power density inside the cavity below 100 mW/cm$^2$ where such effects are expected to be reasonably small. However for measurements at higher power (as in the study of polariton nonlinearities) these issues must be taken into account. 

\begin{figure}[htb]
\centerline{\includegraphics[width=\columnwidth]{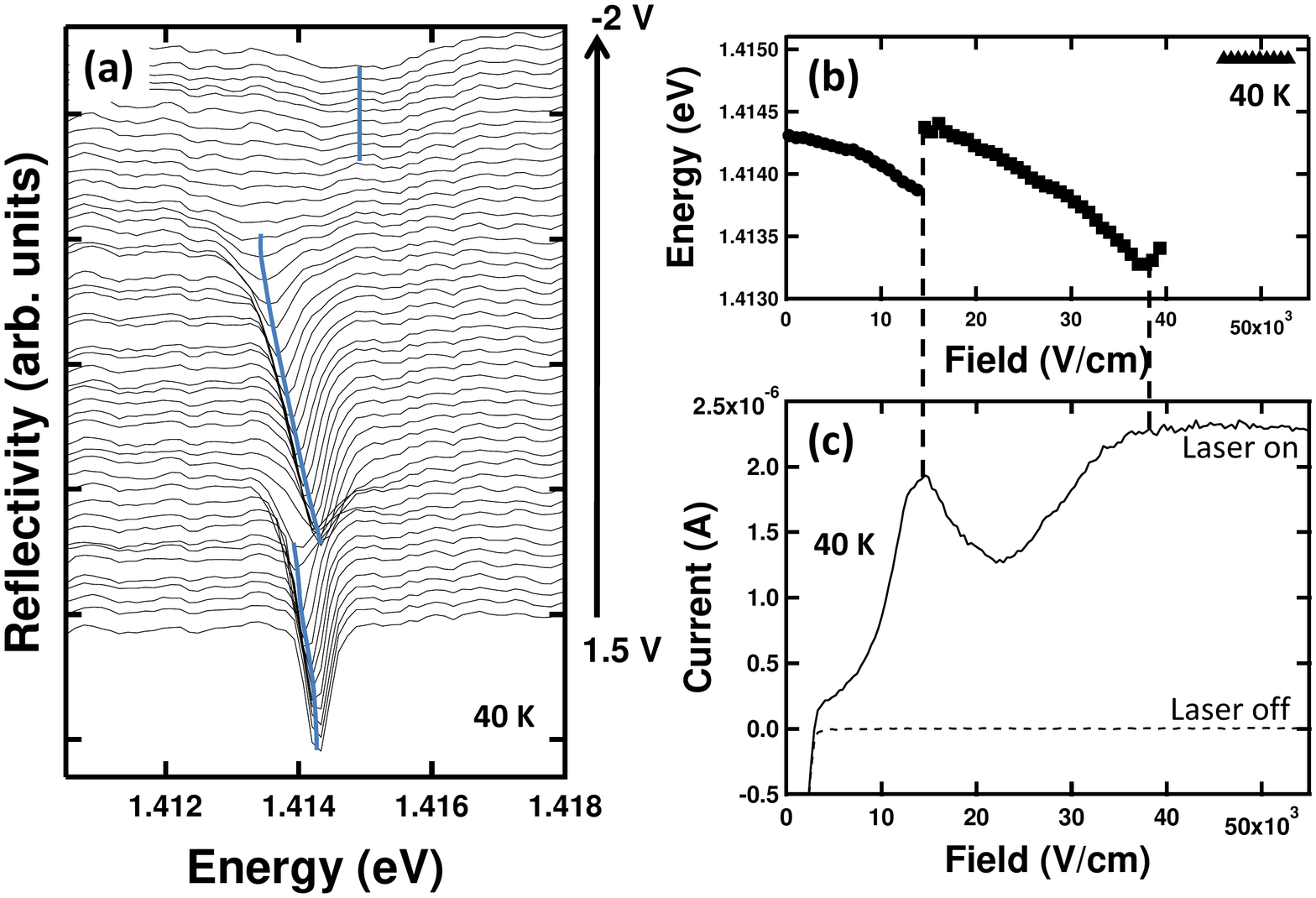}}
\caption{(color online) (a) 40 K reflectivity spectra for bias from 1.5 V to -2 V in steps of 75 mV. Blue (gray) lines are guide to the eye (b) Dispersion curves \textit{vs} internal electric field. (c) Photocurrent \textit{vs} electric field.}
\end{figure}

The extracted dispersion curves with electric field [Fig. 3(b)] show the oriented polariton anticrossing occuring at $14$ kV/cm which is in good agreement with the simulations [Fig. 2(a)] which predict $17$ kV for this detuning. The simulations also explain why the anticrossing jumps rapidly because the photon fraction is very quickly transferred from one branch to the other, making the indirect exciton-like polariton branch very weak. At large electric fields above 37 kV/cm, the features do not track the predicted polariton dispersion. However the simulation shows that at $38$ kV/cm the lifetime of the electron level becomes equal to the Rabi period which means that strong coupling is lost. The non-dispersing mode is therefore the bare cavity  mode. This mode is much broader than at low fields because the QW absorption from the electron-hole continuum has also Stark red-shifted to the same energy at high bias.

The bias-dependent photocurrent corresponding to this tunnelling resonance (Fig.3c) shows a strong peak at 14 kV/cm, and a plateau at higher bias. All photocurrent is created by absorbtion into the polariton modes, as seen by the negligible dark current. Polaritons tunnel out from the RQW with a probability which increases with bias. This probability is increased when direct and indirect polaritons in the LQW first tunnel into the RQW at the oriented polariton resonance. Note that both symmetric and asymmetric tunnelling levels $\left| OP \right\rangle_{\pm}$ are excited by the spectrally broad laser pulses, and the tunnelling rate thus shows a single peak. At higher bias, tunnelling becomes faster than the Rabi period and the photocurrent saturates as all carriers tunnel out before radiative recombination.

The energy splitting of the oriented polaritons produced by the tunnelling allows lateral polariton confinement. Thus gated structures with indirect excitons allow trapping of polaritons in electrostatic traps,\cite{Hammack06} to form waveguides and optical transistors. 

In conclusion we have fabricated and studied an ADQW MC in the strong coupling regime observing both symmetric- and asymmetric-oriented polaritons. The electron QW level coupling is clearly shown experimentally within the LPB and in good agreement with simulations. This effect can be efficiently controlled  by reverse biasing the sample, by tuning the levels into tunnelling resonance. We also show that in such structures a strong dipole moment is created at the oriented polariton resonance which can enhance the polariton-polariton scattering making these structures very promising for studies of nonlinearities. Finally, from a wider perspective, such tunnelling microcavities enable many types of opto-electronic devices. 

The authors acknowledge assistance from N. T. Pelekanos and support from UK EPSRC EP/C511786/1, EP/F011393 and EU CLERMONT4.


%
%

%



\begin{thebibliography}{bib}
\bibitem{noteWF} The levels being only quasiconfined, the wavefunctions are not normalized. As we are interested by the levels confined in the QWs, the integration is stopped when the energy level cross the barrier.
\end{thebibliography}

\end{document}